\documentclass[11pt]{article} 
\usepackage{geometry}
\usepackage{authblk}
\usepackage{amsfonts, amsmath, amsthm, amssymb, enumerate, rotate}
\usepackage{amssymb, bm, bbm, latexsym, mathrsfs}
\usepackage{graphicx}
\usepackage{color, url}
\usepackage{times}
\usepackage{w-thm}
\usepackage{natbib}
\usepackage[title]{appendix}
\usepackage{url}
\usepackage{epstopdf}
\usepackage{paralist} 
\usepackage{float}

\begin{document}

\title{Bayesian Nonparametric Sensitivity Analysis \\of Multiple Test Procedures Under Dependence}
\author[1]{George Karabatsos}
\affil[1]{Departments of Mathematics, Statistics, and Computer Sciences\protect\\
and Educational Statistics \protect
\\ University of Illinois-Chicago \protect\\
e-mail: gkarabatsos1@gmail.com, georgek@uic.edu}
\date{\today}

\maketitle
\begin{abstract}
This article introduces a sensitivity analysis method for Multiple Testing Procedures (MTPs) using marginal $p$-values. The method is based on the Dirichlet process (DP) prior distribution, specified to support the entire space of MTPs, where each MTP controls either the family-wise error rate (FWER) or the false discovery rate (FDR) under arbitrary dependence between $p$-values. This DP MTP sensitivity analysis method provides uncertainty quantification for MTPs, by accounting for uncertainty in the selection of such MTPs and their respective threshold decisions regarding which number of smallest $p$-values are significant discoveries, from a given set of null hypothesis tested, while measuring each $p$-value's probability of significance over the DP prior predictive distribution of this space of all MTPs, and reducing the possible conservativeness of using only one such MTP for multiple testing. The DP MTP sensitivity analysis method is illustrated through the analysis of over twenty-eight thousand $p$-values arising from hypothesis tests performed on a 2022 dataset of a representative sample of three million U.S. high school students observed on 239 variables. They include tests which, respectively, relate variables about the disruption caused by school closures during the COVID-19 pandemic, with various mathematical cognition, academic achievement, and student background variables. \texttt{R} software code for the DP MTP sensitivity analysis method is provided in the Code and Data Supplement of this article.
\newline
\textbf{Keywords:} Dirichlet process; Sensitivity analysis; Multiple testing; False discovery rate; Familywise error rate.
\end{abstract}

\section{Introduction}\label{Section:Introduction}

$P$-values are ubiquitous in science and provide a common ``bottom line" language for statistics communication. Most statistical analyses routinely output dependent (correlated) $p$-values from hypothesis tests, which require using MTPs that are valid under arbitrary dependence between $p$-values. MTPs based on marginal $p$-values remain popular in practice \citep{TamhaneGou18}, despite concerns about them \citep[e.g.,][]{WassersteinLazar16}. This is because marginal $p$-values are easy to apply, are readily available from any statistical software package, and can reduce the results of different (e.g., $t$, Wilcoxon, $\chi^2$, and/or log-rank) test statistics to a common interpretable $p$-value scale, without requiring assumptions or explicit modelling of the potentially-complex joint distributions of the test statistics, having typically unknown correlations. In the modern computing era, the ease at which multiple $p$-values can be computed and interpreted from statistical packages renders problems of multiple inferences and hypothesis testing ubiquitous in the applied sciences, as evidenced by the fact that scientific journals are inundated with $p$-values.

For example, due to the widespread popularity of MTPs based on marginal $p$-values \citep{TamhaneGou22}, the Food and Drug Administration (FDA) extensively discusses them in their Multiplicity Guidance Document on analyzing data from clinical trials \citep[][\S4, Appendix]{USFDA22}, which refers to another document \citep[][]{USFDA21} that discusses the important scientific and public role of sensitivity analysis. Specifically, to analyze the sensitivity of statistical results over a range of deviations from underlying statistical assumptions, in order to examine and enhance the robustness, precision, and understanding of statistical conclusions. Because the $p$-value is a deterministic transformation of a test statistic (e.g., measuring a treatment effect; \citet{Hedges25}) in a standardized way on the $[0,1]$ interval \citep[][p.17]{Dickhaus14}, it is reasonable to apply sensitivity analysis to $p$-values \citep[e.g.,][Ch.4]{Rosenbaum02a} by integrating the outcomes of the sensitivity analysis over explicit prior distributions for parameters that may be driving the sensitivities of these results \citep{Greenland05}. 

This article introduces a sensitivity analysis method for MTPs that are valid under arbitrary dependence between $p$-values, based on a Bayesian non-parametric (BNP), DP prior distribution \citep{Ferguson73}. 

\begin{figure}[!]
\begin{center}
\includegraphics[scale=0.80]{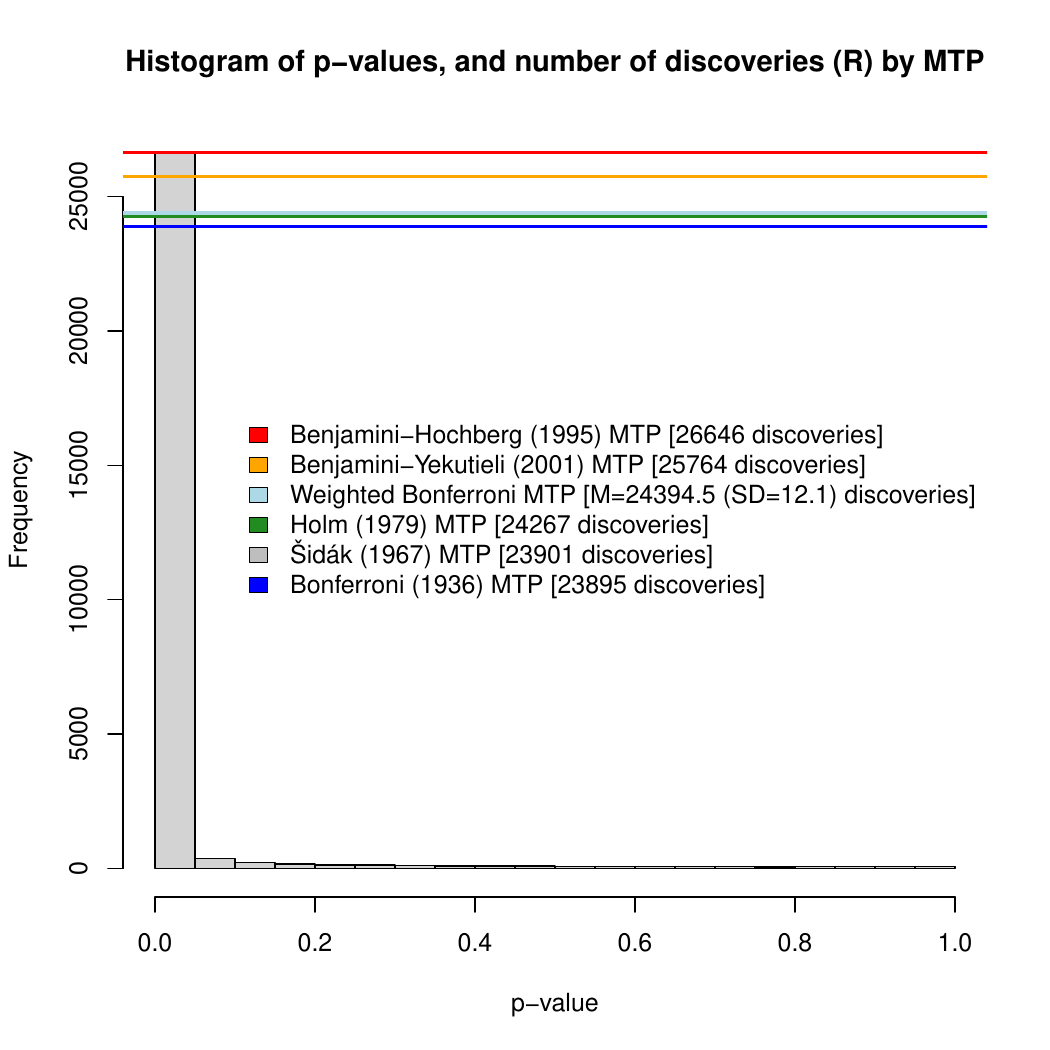}
\caption{A histogram of $p$-values from corresponding $m=28,679$ null hypothesis tests performed on the PISA dataset ($x$). 
Each colored horizontal line refers to an MTP, declaring which number $R_{\alpha}(x)$ (in brackets) of the smallest $p$-values are significant discoveries at the $\alpha = 0.05$ level.}
\label{Figure:pvalsMTPs}
\end{center}
\end{figure}

As a motivation for the proposed sensitivity analysis method, Figure \ref{Figure:pvalsMTPs} presents a histogram of $p$-values from corresponding 28,679 nonparametric (rank-based) two-sided null hypothesis tests performed on 239 key variables of the Programme of International Student Assessment (PISA) 2022 dataset, observed from a representative sample of 3,661,328 age 15 students attending U.S. secondary schools \citep{OECD24}. More details about this dataset are provided in \S\ref{Section:Illustration} and the Appendix. The 28,679 tests include 28,441 tests of zero \citet{Kendall75} partial rank-order $\tau$ correlation on all distinct pairs of the 239 variables, each pairwise correlation obtained after partialling out the effects of the 237 variables \citep[see also][]{Kim15}; and 238 \citet[BM;][]{BrunnerMunzel00} two-sample hypothesis tests of zero gender mean ranking difference, respectively performed on each of the other variables \citep[see also][]{BrunnerEtAl19}. Each of these nonparametric hypothesis testing procedures makes minimal assumptions about the form of the underlying true data-generating distribution (as close as one can get to avoiding all modelling assumptions, \citet[][p.1643]{OHagan13}) while being applicable to dependent variables that are either ordered categorical, count, or continuous-valued; and is based on a permutation distribution of the test statistic that rather quickly converges to a standard normal distribution \citep{Kim15,MaghsoodlooPallos81,NeubertBrunner07}, to enable straightforward corresponding computation of $p$-values. The BM test provides a solution to the Behrens-Fisher problem, unlike the classical Wilcoxon-Mann-Whitney rank sum two-sample test. In any case, the FDA and PIRLS data provide merely two broad examples of the ubiquity of hypothesis testing and $p$-values in the sciences.

Figure \ref{Figure:pvalsMTPs} also presents alternative MTPs, each MTP deciding which number of the smallest $p$-values are significant discoveries, from the 28,679 $p$-values. They include MTPs that conservatively control either the family-wise error rate (FWER) \citep[e.g.,][]{Bonferroni36, Sidak67, Holm79} or the  false discovery rate (FDR) \citep{BenjaminiYekutieli01} under arbitrary dependencies between $p$-values; and an MTP proven to control FDR for independent $p$-values \citep{BenjaminiHochberg95}, and conservatively controls FDR under arbitrary dependence between $p$-values \citep{Farcomeni06} without theoretical guarantees. 

Figure \ref{Figure:pvalsMTPs} reveals that the choice of MTP can noticeably impact the decision regarding which smallest subset of the $m=28,679$ $p$-values are significant discoveries, even though most of these $p$-values are small. The \citet[][]{BenjaminiHochberg95} MTP claims that the 26,646 smallest $p$-values are significant, followed by the \citet[][]{BenjaminiYekutieli01} MTP (25,764), the weighted Bonferroni MTP (claims 24,394.5 $p$-values are significant on average, with standard deviation 12.1, over 1,000 samples of $p$-values weights from the $m$-dimensional $\mathrm{Dirichlet}_m(1,\ldots,1)$ distribution), the \citet[][]{Holm79} MTP (claims 24,267 significant $p$-values), the \citet[][]{Sidak67} MTP (23,901), and the \citet[][]{Bonferroni36} MTP (23,895). Further, selecting one MTP to make decisions about significance does not fully account for the uncertainty in multiple hypothesis testing, while there are many MTPs that a data analyst can choose from. 

The proposed DP MTP sensitivity analysis method addresses these issues, through the specification of a DP prior distribution that supports the entire space of distribution functions, thereby supporting the entire space of MTPs, where each MTP controls either the FWER or the FDR under arbitrary dependencies between $p$-values. Therefore, the DP MTP sensitivity analysis method naturally accounts for the uncertainty in the selection of MTPs and their respective decisions regarding which number of the smallest $p$-values are significant discoveries, from any given set of null hypotheses tested. Also, this method can measure each $p$-value's probability of significance relative to the DP prior predictive distribution of this space of all MTPs. Further, while using any one such MTP can be conservative, the DP prior reduces this conservativeness by supporting a wide range of such MTPs for multiple hypothesis testing. Finally, the DP MTP sensitivity analysis method can be applied to analyze $p$-values arising from any number or combination of different hypothesis testing procedures (not only Brunner-Munzel or rank order correlation tests, but perhaps additional and/or alternative hypothesis testing procedures), provided that for each null hypothesis tested, the $p$-value is super-uniform under the null hypotheses (more details in \S\ref{Section:ReviewMTP}).

The next sections give related background reviews and illustrations to further describe and motivate the DP MTP sensitivity analysis method. Section \S\ref{Section:ReviewMTP} reviews the multiple hypothesis testing framework and traditional MTPs that each controls either the FWER or the FDR under arbitrary dependence between $p$-values. \S\ref{Section:MTP_Sensitivity_Analysis} describes the DP MTP sensitivity analysis method, and then \S\ref{Section:Illustration} illustrates this method through the analysis of the 28,679 $p$-values obtained from the PISA 2022 U.S. dataset. \S\ref{Section:Conclusions} concludes this article and suggests avenues for future research.

\section{Review of MTPs Valid Under Arbitrary Dependence}\label{Section:ReviewMTP} 

To further contextualize and motivate the DP MTP sensitivity analysis method, this section reviews the multiple hypothesis testing and MTPs mentioned in \S1 and some their properties, based on a general theoretical framework of multiple hypothesis testing \citep{BlanchardRoquain08} while referring readers to other articles for related reviews or more technical details.

Let $(\mathcal{X}, \mathfrak{X}, P)$ be a probability space, with $P$ belonging to a set or “model” $\mathcal{P}$ of distributions, while this set can represent a parametric or non-parametric model. A \textit{null hypothesis}, denoted $H$ (or $H_0$), is a subset (submodel) $H\subset\mathcal{P}$ of distributions on $(\mathcal{X},\mathfrak{X})$. And $P\in H$ denotes that $P$ satisfies $H$.

In any application of multiple testing, it is of interest to determine whether $P$ satisfies distinct null hypotheses, belonging to a certain set (family) $\mathcal{H}$ of \textit{candidate null hypotheses}, which is usually a countable set,  or instead can be a continuous (uncountable) set of hypotheses \citep[e.g.,][]{PeronePacificoEtAl04}. From $\mathcal{H}$, let $\mathcal{H}_{0}(P)=\{H\in\mathcal{H}\mid P\in H\}\subseteq\mathcal{H}$ be the set of \textit{true null hypotheses}, and $\mathcal{H}_{1}(P)=\mathcal{H}\backslash\mathcal{H}_{0}(P)$ the set of \textit{(truly) false null hypotheses}, under any (typically unknown) true data-generating distribution $P\in\mathcal{P}$.

In a typical application of multiple hypothesis testing, $\mathcal{H}$ is a finite set of $m$ null hypotheses, given by $\mathcal{H}=\{H_1,\ldots,H_m\}$ with $m=|\mathcal{H}|\in\mathbb{Z^+}$ the number of candidate null hypotheses. Then, there are $m_0(P) = |\mathcal{H}_{0}(P)|\leq m$ \textit{true null hypotheses}, and $m_1(P) = |\mathcal{H}_{1}(P)|=m-m_0\leq m$ \textit{(truly) false null hypotheses}, with $\pi_0(P) = m_0(P) / m$ the \textit{proportion of true nulls}, under any given distribution $P\in\mathcal{P}$.

A \textit{multiple testing procedure (MTP)} is a decision, $\mathcal{R}:x\in\mathcal{X}\mapsto\mathcal{R}(x)\subset\mathcal{H}$, that designates the subset of rejected null hypotheses on the given sampled dataset $x\sim P\in\mathcal{P}$ (with indicator function $\mathbf{1}\{H\in \mathcal{R}(x)\}$ measurable for any $H\in\mathcal{H}$), and $\mathcal{R}^c(x)=\mathcal{H}-\mathcal{R}(x)$ is the subset of non-rejected hypotheses. An MTP $\mathcal{R}(x)$ commits a \textit{Type I error} when it incorrectly rejects a true null hypothesis $H$, i.e., $H\in\mathcal{R}(x)\cap\mathcal{H}_{0}(P)$ on a sampled dataset $x\sim P\in H\subset\mathcal{P}$; and commits a \textit{Type II error} when it fails to reject a false null hypothesis $H$, i.e., $H\notin\mathcal{R}(x)\cap\mathcal{H}_{1}(P)$ on a sampled dataset $x\sim P\in\mathcal{P}\setminus H$.

A typical MTP is a function $\mathcal{R}(\mathbf{p})$ of a family of $p$-values, $\mathbf{p}=(p_{H},H\in\mathcal{H})$. Here, each $p$-value $p_{H}$ (also denoted by $p_i(X)$ or $p_{H_i}(X)$ for $H_i\in\mathcal{H}$ when $\mathcal{H}$ is countable) has the intuitive interpretation that the smaller the $p_{H}$, the more decisively the null hypothesis $H\in\mathcal{H}$ is rejected \citep[e.g.,][p.31]{Efron10LSI}. A $p$-value measures how probable are the observed data, given that the null hypothesis is true; but not the probability of the null hypothesis, conditionally on the observed data \citep[e.g.,][p.19]{Dickhaus14}. Formal definitions of a $p$-value can be found in standard references \citep[e.g.,][Definition 2.1]{Dickhaus14}.

We assume that for each null hypothesis $H\in\mathcal{H}$ there exists a (measurable) \textit{$p$-value function}, $p_{H}:\mathcal{X}\rightarrow\lbrack0,1]$, such that if $H$ is true, then the probability distribution ($\mathbb{P}$) of $p_{H}(X)$ is \textit{marginally super-uniform}:\begin{equation}\label{eq:superUniform}
\mathbb{P}_{X\sim P}[p_{H}(X)\leq t]\leq t,\text{ for }\forall P\in
\mathcal{P},\text{ }\forall H\in\mathcal{H}_{0}(P),\text{ and }\forall t\in
\lbrack0,1].
\end{equation}
The stochastic order assumption (\ref{eq:superUniform}) asserts that the distribution of the $p$-value $p_{H}(X)$ under the null $H\in\mathcal{H}$ is \textit{stochastically lower bounded by a uniform} $\mathcal{U}[0,1]$ \textit{random variable} \citep[][p.966]{BlanchardRoquain08}; i.e., is\textit{ stochastically not smaller than a} $\mathcal{U}[0,1]$ random variate \citep[][p.20]{Dickhaus14}. Indeed, $p$-values are occasionally defined via property (\ref{eq:superUniform}) without making any reference to test statistics or rejection regions \citep[][Definition 8.3.26]{CasellaBerger02}, and without necessarily fully specifying the joint null distribution of the $p$-values of the given set of hypotheses $\mathcal{H}$ being tested \citep{Dickhaus14}.

A $p$-value $p_{H}(X)$ and associated hypothesis testing procedure are said to be \textit{calibrated} if, under the null hypothesis $H\in\mathcal{H}$, the $p$-value has uniform $\mathcal{U}[0,1]$ distribution over random datasets $X\sim P\in H\subset\mathcal{P}$; such a $p$-value satisfies the marginal super-uniform condition (\ref{eq:superUniform}) with strict equality $\mathbb{P}_{X\sim P}[p_{H}(X)\leq t]=t$ for all $t\in\lbrack0,1]$. (Recall that if $U\sim\mathcal{U}[0,1]$, where $\mathcal{U}[0,1]$ is the standard uniform distribution, then $\mathbb{P}(U\leq t)=t$ for all $t\in[0,1]$). A calibrated $p$-value can be interpreted on a universal scale, such as Fisher's scale of evidence for interpreting $p$-values \citep[][p.31, Table 3.1]{Efron10LSI}. Any Neyman-Pearson type test of a simple (e.g., point-null) hypothesis gives rise to a calibrated $p$-value if its underlying test statistic has a stochastically smaller continuous distribution under the null, compared to that under the alternative hypothesis \citep[for more details, see][Theorem 2.2]{Dickhaus14}. For any setting involving $m$ tests of a set of hypotheses $\mathcal{H}$, that gives rise to calibrated $p$-values $p_1,\ldots,p_m$, the joint distribution of these $p$-values under the null hypothesis can be explicitly specified by the so-called independently and identically distributed uniform (i.i.d. $\mathcal{U}[0,1]$) distribution model \citep[][\S2.2.1]{Dickhaus14}.

When the $p$-value $p_{H}(X)$ satisfies the super-uniform condition (\ref{eq:superUniform}) under the null hypothesis $H$, such that the probability inequality is strict for some $t\in \lbrack0,1]$, then there is (canonical first-order) stochastic dominance \citep{QuirkSaposnik62}, such that this $p$-value is stochastically larger than the uniform $\mathcal{U}(0,1)$ random variable, and then the associated null hypothesis testing procedure is \textit{conservative} relative to a calibrated testing procedure. Conversely, if the $p$-value under the null hypothesis $H$ has a stochastically smaller distribution relative to the $\mathcal{U}(0,1)$ distribution, then the associated hypothesis testing procedure is \textit{liberal}, with a higher Type I error rate compared to a calibrated testing procedure. When necessary, one of many techniques can be used to improve the calibration of (or even fully calibrate) the $p$-value, including for a test of a discrete model, composite null hypothesis, or for model checking \citep[e.g.,][and references therein]{DickhausEtAl12, Dickhaus13, HjortEtAl06, Gosselin11, MoranEtAl24}.

Table \ref{tab:Outcomes_m_tests} is a synopsis of key concepts of multiple hypothesis testing, summarizing the possible theoretical outcomes of $m>1$ hypothesis tests \citep[from][Table 18.5]{HastieTibsFriedman09}, for any given dataset $x\sim P\in\mathcal{P}$ randomly sampled from any given (typically unknown) true data-generating distribution $P\in\mathcal{P}$.

In testing a set of hypotheses, $\mathcal{H}$, a traditional criterion for Type I error control is the FWER, which is the probability ($\mathbb{P}$) of making at least one false discovery \citep{HochbergTamhane87,Efron10LSI} under any MTP $\mathcal{R}$, over datasets $X \sim P$ repeatedly sampled from the given true distribution, $P\in\mathcal{P}$, given by:
\begin{equation}\label{eq:FWER}
\mathrm{FWER}_P(\mathcal{R}) = \mathbb{P}_{X\sim P}[\text{Reject any true hypothesis, } H \in\mathcal{H}_{0}(P)\subseteq\mathcal{H}]=\mathbb{P}_{X\sim P}[V(X)\ge1].
\end{equation}
Another criterion is the FDR \citep{BenjaminiHochberg95}, which for any MTP $\mathcal{R}$ is the expected \textit{False Discovery Proportion} (FDP), the expected ($\mathbb{E}$) (average) proportion of false rejections of null hypotheses out of the total number of rejected hypotheses, over many sampled datasets, $X\sim P\in\mathcal{P}$, given by:
\begin{equation}
    \mathrm{FDR}_P(\mathcal{R})=\mathbb{E}_{X\sim P}[\mathrm{FDP}_P(\mathcal{R}(X))]=\mathbb{E}_{X\sim P}\left[\frac{|\mathcal{R}(X)\cap\mathcal{H}_{0}(P)|}{\mathrm{max}\{|\mathcal{R}(X)|,1\}}\right].
\end{equation}

\begin{table}[H]
    \centering
\begin{tabular}
[c]{lccc}
\textbf{Under a true} & \multicolumn{2}{c}{\textbf{Decision, on a sampled data set,} $x\sim
P\in\mathcal{P}$} & \\
\textbf{distribution} $P\in\mathcal{P}$: & Do not reject null $H$ & Reject null $H$ & \\
 & (not significant) & (significant) &
Total\\\cline{2-3}
Null Hypothesis & \multicolumn{1}{|c}{$U(x)=$} &
\multicolumn{1}{|c}{$V(x)=$} & \multicolumn{1}{|c}{}\\
$H\in\mathcal{H}_0(P)$ is true & \multicolumn{1}{|c}{$|\mathcal{R}^{c}(x)\cap\mathcal{H}_{0}(P)|$} &
\multicolumn{1}{|c}{$|\mathcal{R}(x)\cap\mathcal{H}_{0}(P)|$} &
\multicolumn{1}{|c}{$m_{0}$}\\
 & \multicolumn{1}{|c}{\small\textit{\# True Non-discoveries}} &
\multicolumn{1}{|c}{\textit{\# False Discoveries}} &
\multicolumn{1}{|c}{\textit{\# True Null Hypotheses}}\\\cline{2-3}%
Null Hypothesis & \multicolumn{1}{|c}{$T(x)=$} &
\multicolumn{1}{|c}{$S(x)=$} & \multicolumn{1}{|c}{}\\
$H\in\mathcal{H}_0(P)$ is false & \multicolumn{1}{|c}{$|\mathcal{R}^{c}(x)\cap\mathcal{H}_{1}(P)|$} &
\multicolumn{1}{|c}{$|\mathcal{R}(x)\cap\mathcal{H}_{1}(P)|$} &
\multicolumn{1}{|c}{$m_{1}=m-m_{0}$}\\
 & \multicolumn{1}{|c}{\small\textit{\# False Non-discoveries}} &
\multicolumn{1}{|c}{\textit{\# True Discoveries}} &
\multicolumn{1}{|c}{\textit{\# False Null Hypotheses}}\\\cline{2-3}
Total: & $m-R(x)$  & $R(x)=|\mathcal{R}(x)|$ & $m$\\
      & \textit{\# Non-Discoveries} & \textit{\# Discoveries} & \textit{\# Null Hypothesis Tests}\\
      & \textit{(\# Non-Rejections)} & \textit{(\# Rejections)} & 
\end{tabular}
\caption{Possible outcomes of tests of $m$ hypotheses $\mathcal{H}$ in a $2\times2$ classification table, for an MTP $\mathcal{R}$ on a given dataset, $x\sim P$, sampled from a given (typically unknown) true data-generating distribution, $P\in\mathcal{P}$. The \textit{False Discovery Proportion} (FDP) is defined by $\mathrm{FDP}_P(\mathcal{R}(x))=V(x;P)/\mathrm{max}\{R(x),1\}$. \\Over samples of random datasets $X\sim P\in\mathcal{P}$, the \textit{Type I error rate} is $\alpha^{\ast}=\mathbb{E}_{X\sim P}\left[V(X)\right]  /m_{0}$, \\the \textit{Type II error rate} is $\beta^{\ast}=\mathbb{E}_{X\sim P}\left[T(X)\right]  /m_{1}$, and the \textit{power} is $1-\beta^{\ast}=1-\mathbb{E}_{X\sim P}\left[T(X)\right]/m_{1}$.}
\label{tab:Outcomes_m_tests}
\end{table}

Multiple hypothesis testing aims to maximize the expected number of rejections while controlling the FWER or FDR at a preset small level $\alpha\in(0,1)$, typically $\alpha = 0.05$ or $0.01$, etc. Any MTP is said to \textit{strongly control the FWER} (\textit{FDR}, resp.) if $\mathrm{FWER}_P(\mathcal{R})\leq\alpha$ (if $\mathrm{FDR}_P(\mathcal{R})\leq\pi_0\alpha\leq\alpha$, resp.) for any chosen level $\alpha\in(0,1)$ and for all $P\in\mathcal{P}$ (and all $\pi_0\in[0,1]$, resp.) \citep{HochbergTamhane87,BenjaminiHochberg95}. Therefore, under FWER (FDR, resp.) control, any null hypothesis $H_i\in\mathcal{H}$ is rejected if its $p$-value $p_i\leq\alpha$ (if $p_i\leq\pi_0\alpha\leq\alpha$, resp.). Also, $\mathrm{FDR}\leq\mathrm{FWER}$, with $\mathrm{FDR}=\mathrm{FWER}$ if all null hypotheses are true, and therefore $\mathrm{FWER}\leq\alpha$ implies $\mathrm{FDR}\leq\alpha$, meaning that FDR control is more liberal than FWER control, and that a $p$-value declared significant under FWER control is a stronger result than under FDR control. FWER control is typically used in confirmatory studies (e.g., Phase 3 clinical trials) which usually tests a small number (e.g., $\leq20$) of null hypotheses, whereas FDR control is generally used in exploratory (e.g., genomic or microarray) studies, which often involve testing a very large number of null hypotheses (e.g., $m\ge 1000$). FWER control is too stringent and unnecessary for exploratory studies, which only aim to highlight interesting findings \citep{TamhaneGou22}. 

Multiple hypothesis testing can be more powerful when controlling the FDR instead of the FWER. However, the FDR can be manipulated \citep[][\S6]{FinnerRoters01}, such that from a given set of null hypothesis tests, one can increase the chance of rejecting any null hypothesis of interest by artificially adding extremely false null hypotheses that would surely be rejected, to this pool of tests \citep[see also][Chapter 1, \S1.1.2]{CuiEtAl21}. This is a reason why FDR control is not used in clinical trials.

For the PISA dataset, Figure \ref{Figure:pvalsMTPs} compares the behavior of various MTPs that control either the FWER or the FWER. A few of these MTPs are either step-up MTPs or step-down MTPs. To elaborate, let $p_{(1)}\leq\dots\leq p_{(m)}$ be the order statistics of $m$ $p$-values, with corresponding ordered null hypotheses $H_{(1)},\ldots,H_{(m)}$. A \textit{step-up MTP} $\mathcal{R}_{SU}^{\Delta_\alpha}$ specifies a non-decreasing sequence of thresholds, 
$0\leq\Delta_\alpha(H_{(1)})\leq\cdots\leq\Delta_\alpha(H_{(m)})\leq 1$, and then rejects the null hypotheses having the $R_\alpha(x)$ smallest $p$-values, with:
\begin{equation}\label{eq:R.Delta}
  \textit{R}_\alpha(x) = \underset{r\in\{0,1,\ldots,m\}}{\mathrm{max}}\{r : p_{(r)}(x) \leq \Delta_\alpha(H_{(r)})\}.
\end{equation}
where $p_{(0)}\equiv0$. A \textit{step-down MTP} $\mathcal{R}_{SD}^{\Delta_\alpha}$, for step(s) $i=1,2,\ldots\leq m$, rejects $H_{(i)}$ if $p_{(i)}\leq\Delta_\alpha(H_{(i)})$ and then continues to test $H_{(i+1)}$; and otherwise, stops testing without rejecting the remaining hypotheses $H_{(i)},\ldots,H_{(m)}$. 

MTPs that strongly control FWER, under arbitrary dependencies between $p$-values, include the:
\begin{compactitem}
    \item \textit{\citet{Bonferroni36} MTP}, defined by the threshold function $\Delta_{\alpha}(H_{(r)})=\alpha\beta(r)/m=\alpha/m$; 
    \item slightly more powerful \textit{\citet{Holm79} step-down MTP}, defined by $\Delta_{\alpha}(H_{(r)})=\alpha\beta(r)/m=\frac{\alpha}{m-r+1}$;
    \item \textit{weighted Bonferroni MTP}, defined by $\Delta_{\alpha}(H_{(r)})=\alpha\pi(H_{(r)})\beta(r)=\alpha w_{(r)}$, with a weight $w_{i}\in[0,1]$ assigned to each $H_i\in\mathcal{H}$ such that $\sum_{i=1}^{m}w_i=1$ \citep{RubinEtAl06,WassermanRoeder06}, where the Bonferroni MTP assumes equal weights $w_i=1/m$ (for $i=1,\ldots,m$);
    \item \textit{\citet{Sidak67} MTP}, defined by the critical constant $\alpha'= 1 - (1 - \alpha)^{1/m}$ which is slightly larger than the Bonferroni MTP threshold $\alpha/m$, and provides exact FWER control for independent $p$-values, but is conservative (liberal, resp.) when there is positive (negatively, resp.) dependence between $p$-values. 
\end{compactitem}

Strong FWER control of the Bonferroni MTP (unweighted or weighted) follows from his inequality:
\begin{subequations}\label{eq:BonferroniIneq}
\begin{eqnarray}
\mathrm{FWER}_P(\mathcal{R})&=&\mathbb{P}_{X\sim P}\left[\underset{H_i\in\mathcal{H}_0(P)}\bigcup p_{H_i}(X)\leq \alpha w_i\right]\\
&\leq& \underset{H_i\in\mathcal{H}_0(P)}\sum\mathbb{P}_{X\sim P}[p_{H_i}(X)\leq \alpha w_i]=\alpha \underset{H_i\in\mathcal{H}_0(P)}\sum w_i\leq \alpha.
\end{eqnarray}
\end{subequations}
When the $p$-values are highly positively correlated or when $m$ is large, the Bonferroni inequality is not very sharp, making the Bonferroni MTP overly-conservative and to lack power under these conditions. If the $p$-values are independent, then the additive Bonferroni inequality in (\ref{eq:BonferroniIneq}) can be sharpened by choosing the $\alpha w_i$’s to satisfy the multiplicative equality $1-\prod_{i=1}^m(1-\alpha w_i)=\alpha$, from which the \u{S}id\'{a}k MTP threshold $\alpha'$ (see above) equals the common value of the $\alpha w_i$’s. If the $p$-values are positive quadrant dependent \citep{Lehmann66}, then this multiplicative equality changes to an \textit{in}equality with upper-bound $\alpha$, and the \u{S}id\'{a}k MTP becomes conservative \citep[][pp.11-13]{TamhaneGou18}.

The widely used \citet[BH;][]{BenjaminiHochberg95} step-up MTP, which specifies $\Delta_{\alpha}(H_{(r)})=\alpha\beta(r)=\alpha r/m$ with shape function $\beta(r)=r$, was proven to strongly control $\mathrm{FDR}\leq\pi_0\alpha\leq\alpha$ under independent $p$-values, and to conservatively control FDR under positively regression dependent $p$-values \citep{BenjaminiYekutieli01}. The default BH MTP assumes $\pi_0\equiv1$, while the adaptive BH MTP is based on some estimator of $\pi_0$ \citep[e.g.,][and references therein]{BenjaminiEtAl06,MurrayBlume21,NeumannEtAl21,BiswasEtAl22b}. Simulation studies \citep{Farcomeni06, Kim_vandeWiel08} have shown that such a BH MTP can be robust to the types of dependencies among $p$-values that often occur in practice, but it is conservative and provides no theoretical guarantees of FDR control under arbitrary dependence. Also, it is challenging to reliably estimate $\pi_0$ from arbitrarily-dependent $p$-values \citep{BlanchardRoquain09,FithianLei22}.

If $\nu$ is any arbitrary probability measure on $(0,\infty)$, then the step-up procedure with shape function $\beta_{\nu}(r)=\int_0^{r} x \text{d}\nu(x)$ and threshold function $\Delta_{\alpha,\nu}(H_{(r)})=\alpha\beta_{\nu}(r)/m$ for countable hypotheses $\mathcal{H}$ (or threshold function $\Delta_{\alpha,\nu}(H_{(r)})=\alpha\pi(H)\beta_{\nu}(r)$ for countable or continuous hypotheses $\mathcal{H}$) strongly controls $\mathrm{FDR}\leq\alpha\pi_0\leq\alpha$ 
under arbitrary dependencies between $p$-values, where 
$\pi:\mathcal{H}\rightarrow\lbrack0,1]$ is a probability mass function (or probability density function, respectively) with respect to a $\sigma$-finite positive volume measure $\Lambda$ on $\mathcal{H}$, and $\pi_0=\sum_{H\in\mathcal{H}_0}\Lambda(\{H\})\pi(H)$ (or $\pi_0=\int_{H\in\mathcal{H}_0}\Lambda(\{H\})\pi(H)\mathrm{d}H$, resp.) \citep{BlanchardRoquain08,BlanchardRoquain18}. (Here, if the set $\mathcal{H}$ is countable, then $\Lambda$ is a counting measure and $\Lambda(S)=|S|$ is the cardinality of any given set $S\subset\mathcal{H}$; otherwise, if $\mathcal{H}$ is continuous, then $\Lambda$ may be the Lebesgue measure.) Using different weights $\pi(H)$ (or different weights $\Lambda(\{H\})$, respectively) over countable $\mathcal{H}$ gives rise to \textit{weighted $p$-values} (or \textit{weighted FDR}; \citet{BenjaminiHochberg97}, resp.).

For any finite set of unweighted $m$ null hypotheses $\mathcal{H}$ and $p$-values, the \citet[BY;][]{BenjaminiYekutieli01} distribution-free step-up MTP assumes the probability measure $\nu(\{k\}) = (k\sum_{j=1}^m\tfrac{1}{j})^{-1}$ with support in $\{1,\ldots,m\}$, and with corresponding threshold function $\Delta_{\alpha,\nu}(H_{(r)},r)=\alpha\beta_{\nu}(r)/m$ based on linear shape function $\beta_{\nu}(r)=\sum_{k=1}^r k\nu(\{k\})=r/ (\sum_{j=1}^m \tfrac{1}{j})$ \citep[][p.976]{BlanchardRoquain08}. The BY MTP controls FDR for arbitrary dependencies among $p$-values, but it is highly conservative \citep[e.g.,][]{Farcomeni06}. Indeed, other choices of $\nu$ can sometimes improve the power of the corresponding MTP \citep[][\S4.2]{BlanchardRoquain08}. 

Building on this idea regarding power, the next section proposes an MTP sensitivity analysis method, which is based on assigning a distribution on the entire space of the probability measure, $\nu$, in order to induce a distribution of MTP thresholds that are valid under arbitrary dependence among $p$-values.

\section{DP MTP Sensitivity Analysis Method}\label{Section:MTP_Sensitivity_Analysis}

For any probability space $(\mathcal{X},\mathcal{A},G)$, a random probability measure $\nu$ is said to follow a Dirichlet process (DP) prior with baseline probability measure $\nu_0$ and mass parameter $M$, denoted $\nu\sim\mathrm{DP}(M\nu_0)$, if 
\begin{equation}
(\nu(B_1),\ldots,\nu(B_m))\sim\mathrm{Dirichlet}_m(M\nu_0(B_1),\ldots,M\nu_0(B_m))
\end{equation}
for any (pairwise-disjoint) partition $B_1,\ldots,B_m$ of the sample space $\mathcal{X}$ \citep{Ferguson73}. 

The DP prior distribution of $\nu$ has expectation $\mathbb{E}[\nu(\cdot)]=\nu_0(\cdot)$ and variance $\mathbb{V}[\nu(\cdot)]=\frac{\nu_0(\cdot)[1-\nu_0(\cdot)}{M+1}$, and supports the space of discrete random probability measures $\nu$ with probability 1 (almost surely).

Recall from \S\ref{Section:ReviewMTP} that if $\nu$ is any arbitrary probability measure on $(0,\infty)$, then the step-up MTP with shape function $\beta_{\nu}(r)=\int_0^{r} x \text{d}\nu(x)$ strongly controls $\mathrm{FDR}\leq\alpha\pi_0\leq\alpha$ under arbitrary dependence between $p$-values, where $\beta_{\nu}(r)$ corresponds to threshold function $\Delta_{\alpha,\nu}(H_{(r)})=\alpha\beta_{\nu}(r)/m$ that decides which of the smallest $p$ values can be claimed to be significant discoveries, from the $m$ total tests performed of countable hypotheses $\mathcal{H}$ (for example). 

From this perspective, it is possible to assign a DP prior distribution for $\nu$, which in turn induces a prior distribution for the shape parameter $\beta_{\nu}$ and the corresponding threshold parameter $\Delta_{\alpha,\nu}$ and number of discoveries $R_{\alpha}$ from equation (\ref{eq:R.Delta}), thereby treating these functions as random instead of fixed, as done by the standard MTPs. This is the basis of the DP MTP sensitivity analysis method, used to assess and account for uncertainty in the selection of MTPs, and the respective cut-off points and decisions, regarding which of the smallest $p$-values are significant discoveries from a given set $\mathcal{H}$ of hypotheses tested.

Also, the DP MTP sensitivity analysis method measures each $p$-value's probability of significance with respect to the DP prior predictive distribution of all MTPs, each (random) MTP providing either FWER or FDR control under arbitrary dependence between $p$-values. Specifically, for each $p$-value from a set of $m$ ordered $p$-values, $\{p_{(r)}\}_{r=1}^m$, this method counts the proportion of times the $p$-value is significant, with:
\begin{subequations} \label{eq:MTPsens}
\begin{eqnarray}
&&\Pr[\mathbf{1}(r\leq\textit{R}_{\alpha,\nu}(x); p_{(r)})=1, \text{ }r=0,1,\ldots,m]\\
&&=\int\dots\int[\mathbf{1}(r\leq\textit{R}_{\alpha,\nu}(x); p_{(r)})=1,\text{ }r=0,1,\ldots,m]\mathcal{D}_{m}(\mathrm{d}\nu_1,\ldots,\mathrm{d}\nu_m\mid M\nu_{0}),
\end{eqnarray}
\end{subequations}
where 
\begin{subequations}
\begin{eqnarray}
\textit{R}_{\alpha,\nu}(x) &=& \underset{r\in\{0,1,\ldots,m\}}{\mathrm{max}}\{r : p_{(r)}(x) \leq \Delta_{\alpha,\nu}(H_{(r)})\}\\
 &=& \underset{r\in\{0,1,\ldots,m\}}{\mathrm{max}}\left\{r : p_{(r)}(x) \leq \dfrac{\alpha}{m}\sum\limits_{j=1}^{r}j\nu (j-1,j]\right\},
\end{eqnarray}
\end{subequations}
with $p_{(0)}\equiv0$, and where $\mathcal{D}_{m}(\nu_1,\ldots,\nu_m\mid M\nu_{0})$ (with $\nu_r\equiv B_j=\nu(r-1,r]$ for $r=1,\ldots,m$) denotes the cumulative distribution function (CDF) of the $\mathrm{Dirichlet}_m(M\nu_0(B_1),\ldots,M\nu_0(B_m))$ distribution.

Now consider some possible choices of parameters $(M,\nu_0)$ of the DP prior distribution. By default, the DP baseline measure $\nu_0$ can be set to $\nu_0(r-1,r]\equiv (r\sum_{j=1}^m\tfrac{1}{j})^{-1}$, which matches (in prior expectation) the probability measure $\nu$ defining the BY MTP. Also, if there is no further prior information in a multiple testing setting, the DP precision parameter $M$ can be set to a small value, which implies a large prior variance for $\nu$ around the BY baseline $\nu_0$, and thus for $\beta_{\nu}$ and $\Delta_{\alpha,\nu}$. Alternatively, the choice of $M$ may be elicited according to the expected number of clusters under realizations from the DP prior, given by:
\begin{equation}
\mathbb{E}(\text{number of clusters})=\sum_{i=1}^m\frac{M}{M+i-1},
\end{equation}
which takes values in the interval  $[\mathrm{max}\{{1,M\mathrm{log}(\frac{M+m}{M})\}},1+M\mathrm{log}(\frac{M+n+1}{M})]$ \citep[][\S3]{Escobar95}. When $M=m^{-1}$, all $m$ $p$-values are expected to form about one cluster; and when $M=m^2$, they are expected to form about $m$ clusters, where all the $m$ $p$-values are independently and identically distributed (i.i.d.) samples from $\nu_0$. Further, $M$ may be assigned a hyperprior distribution that supports a range of values suitable for applications, while providing MTP inferences that are robust to the choice of $M$. A default hyperprior for $M$ is the standard $\mathrm{Exponential}(1)$ distribution, because it supports a wide range of $\nu$ and the expected number of clusters, while mostly supporting low values of $M$, each of which implies a large prior variance for $\nu$ and expects a few clusters (groups) of $p$-values.

The DP MTP sensitivity analysis method emphasizes prior predictive inference as in \citet{Box80}, and adopts the view that for statistical analysis, the Bayesian ingredients of likelihood and prior should be kept separate while making their full background information explicitly available \citep[][p.314; p.7; resp.]{Fraser11,FraserReid16}, instead of combining them into a posterior distribution. Previous BNP models for multiple testing focus inference on the posterior distribution of model parameters, including DP mixture models for the $p$-value distribution \citep[e.g.,][]{TangEtAl07,GhosalEtAl08, WangGhosal25} and for the summary statistics distribution \citep{DunsonEtAl08, GhosaRoy11, DentiEtAl20}.

The DP MTP sensitivity analysis method can be applied to analyze any given set of $p$-values from multiple hypothesis tests (based on Monte Carlo samples from the prior predictive distribution (\ref{eq:MTPsens}), integrated over the hyperprior distribution), by using a few lines of basic \texttt{R} software code \citep{RCoreTeam24}, contained in the file \texttt{DP-MTP.R} available in the Code and Data Supplement section of this article.

\section{Illustration of DP MTP Sensitivity Analysis}\label{Section:Illustration}

Recall Figure \ref{Figure:pvalsMTPs} (in \S\ref{Section:Introduction}), which shows the histogram of the two-tailed $p$-values from 28,679 hypothesis tests computed on 239 variables of the PISA 2022 U.S. dataset, using the Kendall's (1975\nocite{Kendall75}) partial rank-order $\tau$ correlation and \citet{BrunnerMunzel00} test procedures.

The PISA 2022 dataset was observed from a representative sample of 3,661,328 age 15 students attending U.S. secondary schools \citep{OECD24}. These variables describe individual students': home, family, and teacher and school background; learning experiences, attitudes, dispositions, beliefs, and well-being; and the teaching practices, learning environment, and perceptions and involvement, provided by the students' respective teachers, school principals and organization, and family; and students' individual scores in math, reading, science, and in eight specialized math subdomains; and other variables individually describing the students, and their respective parents, teachers, classrooms, and schools. (See the Appendix for more details). Variables include students' individual responses on a 45-item rating scale questionnaire (either 2-, 3-, mostly 4-, or 5-point rating items) asking on how much school closures caused by the COVID-19 pandemic disrupted their lives and impacted their access to various teaching and learning practices, and to other school resources \citep{OECD23}; and continuous variables on fostering reasoning and on encouraging mathematical thinking, representing students' cognitive activation in mathematics. In summary, this large-scale hypothesis testing analysis of this PISA dataset, among other things, aims to analyze the partial correlation between COVID-19 impacts, cognitive activation in math, and student performance in math, reading, and science; and to explore gender differences on each of the other variables. Accordingly, the null hypothesis tests include 28,441 tests of zero Kendall's (1975) partial rank-order $\tau$ correlation \nocite{Kendall75} on distinct pairs of the 239 variables; and 238 \citet{BrunnerMunzel00} two-sample hypothesis tests of no gender difference in mean ranking, on each of the other variables. (The Code and Data Supplement includes an \texttt{R} code file that describes how to compute each of these testing procedures based on the PISA sampling weights).

\begin{figure}[!]
\begin{center}
\includegraphics[scale=0.88]{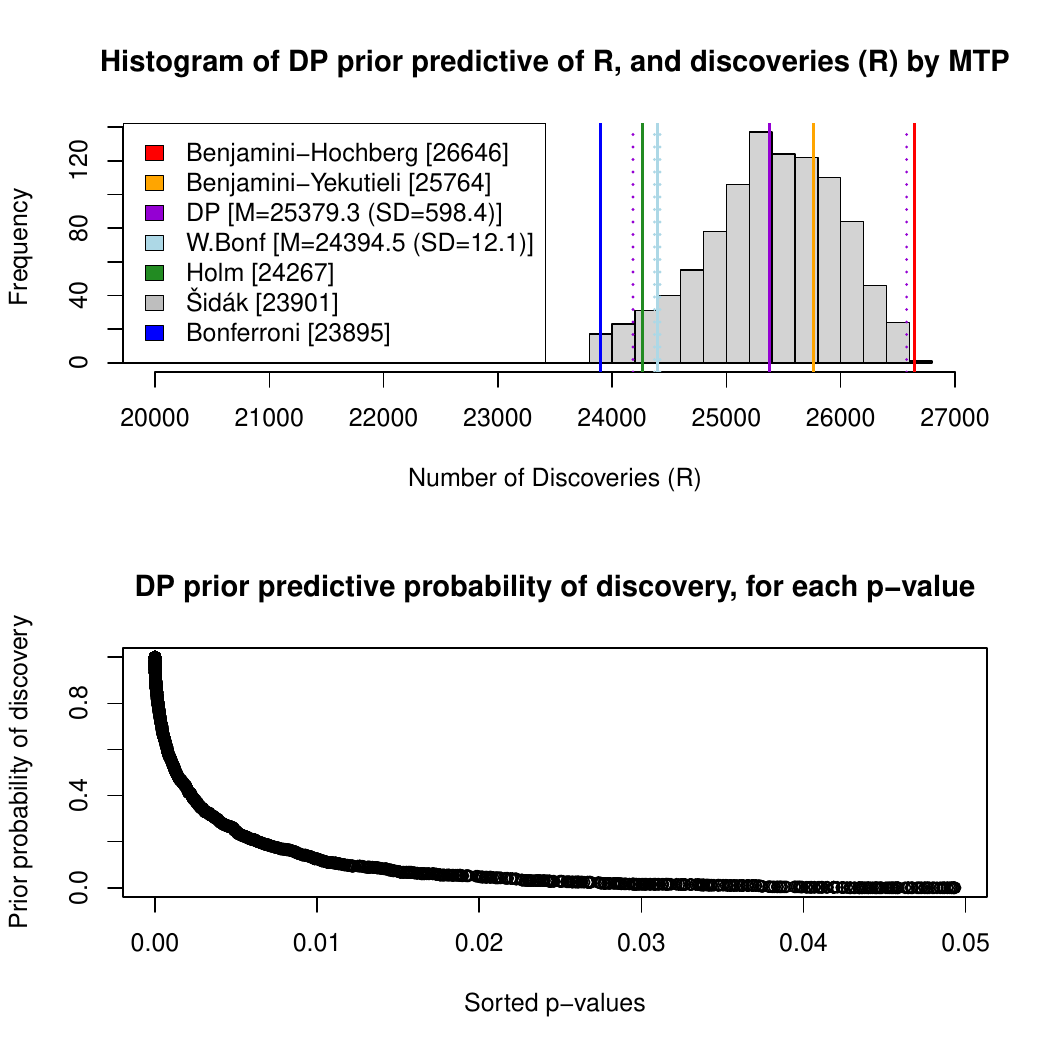}
\caption{
\textbf{Top panel:} 
For the analysis of $p$-values obtained from the corresponding $m=28,679$ null hypothesis tests performed on the PISA dataset ($x$), a histogram of the DP prior predictive distribution of the number of discoveries, $R_{\alpha}(x)$, based on 1,000 samples $(\nu(B_1),\ldots,M\nu(B_m))\sim\mathrm{Dirichlet}_m(M\nu_0(B_1),\ldots,M\nu_0(B_m))$ of the random probability measure $\nu$ drawn from the DP prior distribution, with baseline measure $\nu_0$ specified by BY MTP, and corresponding samples from the $M\sim\mathrm{Exponential}(1)$ hyperprior. 
Each colored vertical line refers to an MTP, declaring which number $R_{\alpha}(x)$ (in brackets) of the smallest $p$-values are significant discoveries at the $\alpha = 0.05$ level, as mentioned in Figure \ref{Figure:pvalsMTPs}, \S1, and \S4.
\textbf{Bottom panel:} The DP prior predictive probability of significance (discovery), for the subset of $p$-values $\leq \alpha=0.05$, based on the 1,000 DP samples of the random measure, $\nu$.}
\label{Figure:MTP_Sensitivity_Analysis}
\end{center}
\end{figure}

For the analysis of the $m=28,679$ $p$-values obtained from the corresponding null hypothesis tests performed on the PISA dataset ($x$), Figure \ref{Figure:MTP_Sensitivity_Analysis} presents the results of the DP MTP sensitivity analysis method, described in \S\ref{Section:MTP_Sensitivity_Analysis}, after applying the \texttt{R} code for the DP MTP method to analyze these $p$-values. 

The top panel of Figure \ref{Figure:MTP_Sensitivity_Analysis} presents the histogram of the DP prior predictive distribution of the number of discoveries (significant $p$-values), $R_{\alpha}(x)$, based on 1,000 samples, $(\nu(B_1),\ldots,M\nu(B_m))\sim\\\mathrm{Dirichlet}_m(M\nu_0(B_1),\ldots,M\nu_0(B_m))$, of the random probability measure $\nu$ drawn from the DP prior distribution, with baseline measure $\nu_0$ specified by BY MTP, and corresponding samples from the $M\sim\mathrm{Exponential}(1)$ hyperprior distribution, while controlling for either the FWER or FDR at the $\alpha = 0.05$ level, under arbitrary dependencies between the $p$-values.

This (histogram) prior predictive distribution of $R_{\alpha}(x)$ expresses the uncertainty quantification of the MTP induced by the DP, i.e., by the DP-MTP process. In particular, the top panel of Figure \ref{Figure:MTP_Sensitivity_Analysis} shows that using the random DP MTP thresholds $\Delta_{\alpha,\nu}$ (and corresponding random $R_{\alpha}(x)$) collectively leads to a greater number and range of $p$-values being declared as significant. On average, the DP MTP declares 25,379 of the 28,679 $p$-values as significant discoveries, with standard deviation 598, and this average falls between the number of discoveries claimed by the Benjamini-Yekutieli (BY) MTP (25,764) and the average number of discoveries of the weighted Bonferroni MTP (24,394.5).

This histogram of the DP prior predictive distribution of $R_{\alpha}(x)$ overlaps with the number of significant $p$-values (discoveries) $R_{\alpha}(x)$ claimed by all the other MTPs (indicated by vertical lines in the figure) that control either the FWER or the FDR under arbitrary dependencies between the $p$-values. This overlap implies that the DP MTP method can control the FWER and the FDR, and in particular, the overlap with the Benjamini-Hochberg (BH) MTP supports the previous claim that this MTP can control FDR under dependence \citep{Farcomeni06}. Also, using any single one of these other MTPs (thresholds), which are valid under arbitrary dependence between the $p$-values, can often lead to conservative decisions as to which number of the smallest $p$-values are significant discoveries, because these other MTPs do not provide uncertainty quantification about $R$. This is true especially for any MTP ($R_{\alpha}$ values) located towards the left side of the graph in the top panel of Figure \ref{Figure:MTP_Sensitivity_Analysis}.

The bottom panel of Figure \ref{Figure:MTP_Sensitivity_Analysis} shows the corresponding DP prior predictive probability of significance, for the subset of the 28,679 $p$-values that are $\leq 0.05$. Clearly, the prior predictive probability of significance increases as the $p$-value decreases, leading to greater confidence of a significant result. The Code and Data Supplements (CDS) section of this article provides an output table with details about which partial Kendall's $\tau$ and Brunner-Munzel (BM) hypothesis tests are likely to be significant based on this prior predictive distribution. All BM tests were significant. This CDS section also provides \texttt{R} code with instructions that can be used to reproduce the results in Figures \ref{Figure:pvalsMTPs}-\ref{Figure:MTP_Sensitivity_Analysis} and in this output table.

Finally, both panels of Figure \ref{Figure:MTP_Sensitivity_Analysis} shows that the DP MTP sensitivity analysis method, while providing uncertainty quantification of MTPs, also provides an ensemble method and a new Vibrations of Effects (VoE) analysis for MTPs. VoE \citep[see][and references therein]{VinatierEtAl25} is a generalized sensitivity analysis method which considers and explores the range of results that can arise from analytical flexibility in which all uncertain analytical and methodological choices are systematically varied to estimate how much different results can be. While previous presentations of the VoE method focus on how results can vary from multiple tests of a \textit{single} hypothesis, the DP MTP method generalizes this by providing a VoE analysis (for MTPs) based on tests of multiple hypotheses.

\subsection{Inference from the Posterior Distribution of $M$}

As an aside (in response to a reviewer's request), the analyst may infer from the posterior distribution of the DP precision parameter $M$, which is given by $\pi(M\mid k)\propto\pi(M)M^{k-1}(\alpha+m)\int_0^1 z^\alpha(1-z)^mdz$ up to a proportionality constant, conditionally on the (fixed) observed $k=28,679$ distinct $p$-values and the $\mathrm{Exponential}(1)$ prior density function, $\pi(M)=\mathrm{exp}(-M)$. (Such a large number of distinct $p$-values is common in settings of large-scale multiple testing, owing to the fact that they usually correspond to continuous-valued test statistics, some which may be obtained after continuing any available discrete-valued test statistics to achieve calibrated $p$-values, as mentioned in \S2). The posterior distribution of $M$ can be estimated using a simple Gibbs MCMC sampler, which generates a new sample of $M$ (given the fixed $k$ and the previous state of $M$) from a simple mixture of two gamma distributions, in each sampling iteration \citep[][pp.584-585]{EscobarWest95}.

From 10,000 converged Gibbs samples of $M$ (obtained after generating 10,000 burn-in Gibbs samples), it was found that $M$ had estimated posterior mean 13364.73, standard deviation 95.58, and 5-number summary quantiles 0\% = 12969.02, 25\% = 13299.80, 50\% = 13364.95, 75\% = 13428.71, 100\% = 13725.31. Correspondingly, the maximum (over $r=1,\ldots,m$) of the variance $\mathbb{V}[\nu((r-1,r])]=\frac{\nu_0((-r1,r])[1-\nu_0((r-1,r])}{M+1}$ of the random probability measure, $\nu$, (recall that DP baseline measure was specified as $\nu_0(r-1,r]\equiv (r\sum_{j=1}^m\tfrac{1}{j})^{-1}$, corresponding to the BY MTP), has posterior mean 0.00001, standard deviation 0.00000, and 5-number summary quantiles all equaling 0.00001, with respect to the posterior of $M$. This choice of $\nu_0$, and this small posterior variance of $M$, both imply that the DP MTP based on the posterior distribution of $M$ essentially corresponds to inference from the Benjamini-Yekutieli MTP (already shown in Figure \ref{Figure:MTP_Sensitivity_Analysis}), which is considered to be overly-conservative by the general MTP literature. (The CDS section also provides \texttt{R} code that reproduces these results from the posterior of $M$).

Hence, we reemphasize prior predictive inference, instead of inference from the posterior distribution of $M$. This is because this prior predictive is based on a DP prior supporting a wider range of the random probability measure $\nu$, and corresponding MTPs that are valid under arbitrary dependence between the $p$-values, including more powerful and less conservative MTPs.

\section{Conclusions}\label{Section:Conclusions}

Scientific fields frequently apply multiple hypothesis testing and report corresponding (marginal) $p$-values, while typical settings of multiple hypothesis involve correlated $p$-values. Meanwhile, any one of the available (cited) MTPs can conservatively control either the FWER or FDR under arbitrary dependence between $p$-values. Also, selecting one of the MTPs for use in multiple hypothesis testing does not fully account for the uncertainty in making decisions about which number of the smallest $p$-values can be declared as significant, from any given set of null hypothesis tests.

The DP MTP sensitivity analysis method addresses these issues by enabling the data analyst to assess and account for uncertainty in the selection of MTPs, and their corresponding decisions regarding which of the smallest $p$-values are significant discoveries from the given set of hypothesis tests. This method achieves this by basing multiple hypothesis testing on the DP prior distribution, specified to support the space of MTPs, each MTP controlling either the FWER or FDR under arbitrary dependence between $p$-values. Also, from $p$-values obtained from a set of hypothesis tests (resp.), the method measures each $p$-value's probability of significance relative to the DP prior predictive distribution of all MTPs that are each valid under arbitrary dependence. Finally, this method can be routinely applied for datasets involving a very large number of hypothesis tests, as illustrated by the analysis of the large PISA 2022 dataset.

The DP MTP sensitivity analysis method can be easily extended to handle (un)weighted countable or continuous hypotheses $\mathcal{H}$ and/or to (un)weighted $p$-values, after making straightforward adjustments to this method. Also, the DP MTP method can be easily extended to handle tests of continuous hypotheses, either by still treating $\nu$ as a discrete random probability measure, and thus assigning it a DP prior; or by viewing $\nu$ as continuous, and thus assigned it a BNP prior distribution with support on the space of continuous random probability measures, such as a DP mixture of continuous densities \citep{Lo84}. 

In future research, the DP MTP sensitivity analysis method can be extended through the use of BNP priors beyond the DP \citep[e.g.,][]{LijoiPrunster10}; and extended to handle online multiple hypothesis testing over time with FDR and FWER control, say, based on the number of discoveries made so far \citep[e.g.,][]{JavanmardMontanari15} and with corresponding test levels $\alpha_1,\alpha_2,\ldots$ that sum to the specified $\alpha$.

\section*{Acknowledgments}
This research is supported in part by Spencer Foundation grant SG200100020; National Science Foundation grants SES-0242030 and SES-1156372; and National Institute for Health grants R01HS1018601 and 1R01AA028483-01. The author thanks an anonymous Reviewer, Associate Editor, and the Editor for providing editorial suggestions that improved the presentation of this article. This article was presented at the 14th International Conference on Bayesian Nonparametrics (BNP14) held in UCLA on Jun 26, 2025. The first version of this article was made available as an \texttt{arXiv} e-print \texttt{arXiv:2410.08080} on October 10, 2024. The author declares no conflict of interest.

\appendix
\begin{center}
\Large\textbf{Appendix: More Details on the PISA 2022 Dataset}
\end{center}

Since 2000, the Organization for Economic Cooperation and Development (OECD) via the Programme for International Student Assessment (PISA) triennially conducts a worldwide survey of age 15 and Grade 10 students, aside from a one-year delay due to the COVID-19 pandemic. Each PISA survey assesses how much students learned from school the essential knowledge and skills in reading, mathematics, and science needed to fully participate in modern societies and address real-life challenges. Specifically, PISA assesses whether students can reproduce, extrapolate from, and apply learned knowledge to new situations, while emphasizing mastery and understanding of processes and concepts, and the ability to function in many situations. Each PISA survey collects data using rigorous technical standards and best educational assessment practices to inform national and international education policy decisions.

The PISA 2022 student assessment emphasized mathematics more than the reading and science domains; and added creative thinking as an innovative domain, measured by two continuous cognitive activation in math variables, namely, fostering reasoning and encouraging mathematical thinking \citep{BurgeEtAl15}. PISA also administered: student questionnaires, asking questions about them and their attitudes, beliefs, and dispositions, and about various aspects of their home, family and school background (e.g., index of economic, cultural and social status (ESCS); \citet{Avvisati20}) and learning experiences, and about their financial literacy; optional student questionnaires on familiarity with information and communications technology and computers, and on well-being; school principal questionnaires on various aspects of school management and organization, and educational provision in schools, and learning environment; and the students’ parent and teacher questionnaires, asking parents about their perceptions of and involvement in their child’s school and learning, and asking teachers about themselves and their teaching practices. Students and school principals in their respective questionnaires also answered questions from the PISA 2022 Global Crises Module \citep{BertlingEtAl20}, which assesses aspects of the disruption caused by the school closures during the COVID-19 pandemic to students, including how these closures affected student lives and school policies, and the measures taken by schools to address this disruption.

The worldwide PISA 2022 student questionnaire dataset was downloaded from \url{https://survey.oecd.org/index.php?r=survey/index&sid=197663&lang=en}. A data subset on 4,552 U.S. students (typically) age 15 and grade 10 U.S. students from 154 U.S. secondary schools, observed on 243 variables, is provided in Supplementary Information. The variables include two variables on country and school and student identification number (variables named \texttt{CNTSCHID} and \texttt{CNTSTUID}, resp.), two student observation weight variables (\texttt{W\_FSTUWT} and \texttt{SENWT}), and 239 survey and assessment variables of the type described in the previous paragraph, with each missing student observation recoded to an empty value. The PISA 2022 U.S. dataset provides a representative sample of $n_w=3,661,328.4945$ students of age 15 and grade 10, with effective sample size $n_w$ equal to the sum of the PISA 2022 Final Trimmed Nonresponse Adjusted Student Weights (\texttt{W\_FSTUWT}), $\{w_i\in \mathbb{R}^+\}_{i=1}^{n=4,552}$, summed over the 4,552 U.S. students, who were obtained by a two-stage stratified sampling design \citep[][Ch.6]{OECD24}. The first stage randomly sampled 154 U.S. secondary schools with probability proportional to the number of eligible school-enrolled age 15 students. The second stage, from each selected school, randomly selected 42 eligible students, or otherwise selected all the available 42 or less eligible students.

Statistical analyses of the PISA 2022 dataset should be based on student survey weights, because these weights scale up the size of the student sample to the size of the U.S. population, while adjusting for student and school over- and undersampling, characteristics, and survey non-response \citep[][Ch. 10]{OECD24}. The sampling observation weight, $w_i$, for each student $i$ equals the product of the student's school base weight (inverse probability that school is randomly selected), the within-school base weight (inverse probability of selecting this student from within the selected school), and four adjustment factors. The first two adjustment factors respectively compensate for non-participation by other schools that are similar to the student's school, if necessary; and for non-participation by students within the same school non-response and explicit stratum categories, and within the same grade and gender categories if the student's school sample size is sufficiently large. The last two adjustment factors, respectively, use trimming to reduce unexpectedly large values of the school base weights; and to reduce the weights of students with exceptionally large values for the product of all the other weight components \citep[][Ch. 10]{OECD24}.

The 28,679 Kendall's partial $\tau$ rank correlation and Brunner-Munzel hypothesis tests conducted on the PISA 2022 U.S. dataset were computed based on these PISA 2022 survey weights, in order to produce statistical results that represent the U.S. age 15 student population. (For more details, see the Code and Data Supplement,  which includes an \texttt{R} code files that describes how to compute each of these testing procedures based on the PISA sampling weights). For any given dataset containing $n$ data points, these weights are denoted $\{w_i\in \mathbb{R}^+\}_{i=1}^{n}$ (e.g., PISA weights), and the effective sample size of the dataset is $n_w=\sum_{i=1}^n w_i$. For an unweighted dataset of $n$ observations, $\{w_i=1\}_{i=1}^{n}$ and $n_w=n$.

The PISA 2022 U.S. dataset contains assessed student achievement scores in mathematics, reading, and science, and in 8 math subdomains: change and relationships; quantity; space and shape; uncertainty and data; employing mathematical concepts, facts, and procedures; formulating situations mathematically; interpreting, applying, and evaluating mathematical outcomes; and reasoning. For each U.S. student, and each of these 11 total domains, the dataset contains 10 plausible values of student domain achievement scores, which are multiple imputation samples \citep{Rubin87} from a posterior distribution of the ability parameter (treated as missing data) of the multiple-group IRT model \citep[for binary and/or polytomous-scored items;][]{BockZimowski97,vonDavierYamamoto04} used to account for measurement error uncertainty arising from the fact that the student was administered and answered a subset of test questions from the full assessment battery \citep[][Ch. 10]{Wu05,OECD24}.

The 28,679 nonparametric hypothesis tests performed on the PISA 2022 U.S. student dataset include tests involving at least one of the 10 plausible values from the 11 assessment domains. But any suitable MTP can be used to combine the results of $p$-values \citep[][\S2.1]{CinarViechtbauer22} arising from hypothesis tests performed on the 10 plausible values (variables) on the given assessment domain, for each of the 11 domains. This implies that the DP MTP sensitivity analysis method can naturally handle the analysis of plausible values. For example, given a number $q$ of $p$-values, one method \citep{Tippett31} calculates the combined $p$-value as $p_c = 1-(1-\mathrm{min}(p_1,\ldots,p_q))^q$, which is the \citet{Dunn58}-\citet{Sidak67} correction based on the \citet{Sidak67} MTP.

\bibliographystyle{apalike2}
\bibliography{Karabatsos.bib}

\end{document}